\documentclass[letterpaper, 10pt, conference]{ieeeconf}
\IEEEoverridecommandlockouts
\newif\ifarxivversion

\arxivversiontrue 

\usepackage{amssymb}
\usepackage{amsmath,amsfonts}
\usepackage{cite}
\usepackage{graphicx}
\usepackage{hyperref}
\usepackage{amssymb}
\usepackage{mathrsfs}
\usepackage{array}
\usepackage{mathabx}
\usepackage{xcolor}

\usepackage{amsthm}
\newtheorem{theorem}{Theorem}
\newtheorem{lemma}{Lemma}

\newtheorem{corollary}{Corollary}

\def\Phibf{\mathbf{\Phi}}
\def\Psibf{\mathbf{\Psi}}

\def\RRational{\frac{1}{s}\mathcal{RH}_\infty}
\def\H2{\mathcal{H}_2}
\def\Hinf{\mathcal{H}_\infty}
\newcommand{\norm}[1]{\left\| #1 \right\|}

\def\Acal{\mathcal{A}}
\def\Bcal{\mathcal{B}}
\def\Ccal{\mathcal{C}}

\def\D{\mathbb{D}}
\def\Pcal{\mathcal{P}}
\def\Qcal{\mathcal{Q}}
\def\Zcal{\mathcal{Z}}

\def\R{\mathbb{R}}

\title{\LARGE \bf Distributed Continuous-Time Control via System Level Synthesis}

\author{Yaozhi Du, Jing Shuang (Lisa) Li
\thanks{The authors are with the Department of Electrical Engineering and Computer Science, University of Michigan.  
 {\tt\small \{yzdu, jslisali\}@umich.edu}}
}
  
\begin{document}

\maketitle

\begin{abstract}

This paper designs $\H2$ and $\Hinf$ distributed controllers with local communication and local disturbance rejection. We propose a two-step procedure: first, select closed-loop poles; then, optimize over parameterized controllers. We build on the system level synthesis (SLS) parameterization --- primarily used in the discrete-time setting --- and extend it to the general continuous-time setting. We verify our approach in simulation on a 9-node grid governed by linearized swing equations, where our distributed controllers achieve performance comparable to that of optimal centralized controllers while facilitating local disturbance rejection.

\end{abstract}

\section{Introduction}
\allowdisplaybreaks

When regulating the dynamics of large-scale systems (e.g., power grids, process plants), the use of centralized controllers is inefficient due to communication and computational requirements. Distributed controllers offer a more practical alternative; specifically, in scenarios where networked systems which are controlled by networks of sub-controllers, \textit{local communication} and \textit{local disturbance rejection} are desirable properties 
\cite{anderson2019system}. By local communication, we mean that sub-controllers communicate only with nearby sub-controllers; by local disturbance rejection, we mean that disturbance effects are confined to local neighborhoods. Rigorous definitions are provided in Section \ref{sec:overview}.

Controllers with local communication can be designed using techniques from structured control 
\cite{sabuau2023network}; however, additional assumptions (e.g., decentralized plants) are often required.
Local disturbance rejection is typically induced through tuning or analyzed post-hoc \cite{motee2017sparsity}; it is rarely enforced (i.e., via constraints).
To the best of our knowledge, no prior work provides a design method for general continuous-time plants that simultaneously accommodates local communication and explicitly enforces disturbance rejection --- the goal of this paper is to propose such a method. We leverage the system level synthesis (SLS) parameterization \cite{anderson2019system} to do so.

The SLS method parameterizes structured controllers using their closed-loop responses. This allows the incorporation of both local communication and local disturbance rejection via sparsity constraints; the resulting problem is convex. Nearly all results on SLS are in the discrete-time setting. The only exception is \cite{jensen2021explicit}, which is restricted to spatially-invariant and spatially-varying plants with decoupled dynamics.

We first provide the problem setup (Section \ref{sec:overview}) and port the SLS parameterization to continuous time (Section \ref{sec:sls}). We then propose a two-step design procedure: first, select closed-loop poles; then, optimize over closed-loop responses. For the first step, we use simple pole approximation (Section \ref{sec:SPA_technique}); for the second step, we formulate state feedback $\mathcal{H}_2$ and $\mathcal{H}_\infty$ control as convex optimizations (Section \ref{sec:optimal_phi}), providing both LMI- and non-LMI based formulations. Finally, we validate our approach in simulations (Section \ref{sec:Simulation}).

\textbf{Notation:} Bold upper-case letters (e.g., $\mathbf{\Phi}(s)$) denote matrices of transfer functions; bold lower case letters (e.g., $\mathbf{u}(s)$) denote signals in $s$ domain. $\mathcal{RH}_\infty$ is the Hardy space of real, rational proper transfer functions. $\Phibf(s) \in \RRational$ iff $s\Phibf(s) \in \mathcal{RH}_\infty$. For matrix $A$, $A^\top$, $A^*$ and $\overline{A}$ represent its transpose, complex-conjugate transpose, and complex conjugate, respectively. $\mathbf{0}_n \in \mathbb{R}^n$ is a zero vector; $I_n \in \mathbb{R}^{n \times n}$ is an identity matrix; $\D$ represents the open unit disk.

\section{Problem setup} \label{sec:overview}
\subsection{Local communication and control} \label{sec:problem_setup}
Consider the linear time invariant (LTI) system:
\begin{equation}
    \dot{x}(t) = A x(t) + B u(t) + w(t) 
    \label{equ:LTIsys}
\end{equation}
where $x(t)\in\R^n$, $u(t)\in\R^m$, and $w(t) \in \R^n$ represent state, control action, and exogenous disturbance. The system contains $N$ interconnected subsystems, each having one or more states. State, control, and disturbance are partitioned into $[x]_i$, $[u]_i$, and $[w]_i$ for each subsystem $i$; system matrices $A$ and $B$ are partitioned into $[A]_{ij}$, $[B]_{ij}$, which capture subsystem-level interaction. System topology is described by an unweighted directed graph $\mathcal{G}(\mathcal{V}, \mathcal{E})$, where vertex $v_i$ corresponds to subsystem $i$, and edge $e_{ij} \in \mathcal{E}$ exists whenever $[A]_{ij} \neq 0$ or $[B]_{ij} \neq 0$. Let $\mathcal{A}\in \{0,1\}^{N \times N}$ be the adjacency matrix for this graph, where by convention $\mathcal{A}_{ii} = 1$.

We seek a state feedback controller of the form 
\begin{equation} \label{eq:controller_time_domain}
    u(t) = \mathcal{K}(x(0:t))
\end{equation}
where $\mathcal{K}$ is linear and causal. Equivalently, in signal domain: 
\begin{equation}
    \mathbf{u}(s) = \mathbf{K}(s)\mathbf{x}(s) \label{equ:state_feedback}
\end{equation}
where $\mathbf{K}(s)$ can be implemented as a distributed controller by partitioning along subsystems; row $[\mathbf{K}(s)]_{i,;}$ is the sub-controller at subsystem $i$, and $[\mathbf{K}(s)]_{ij}$ represents information required by sub-controller $i$ from sub-controller $j$.

\begin{figure} 
 \begin{minipage}{0.64\linewidth}
    \includegraphics[width=\linewidth, page=2]{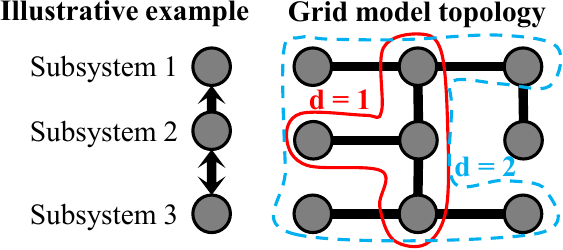}
\end{minipage}
\begin{minipage}{0.35\linewidth}
    \caption{(Left) Graph topology for \eqref{eq:example}. (Right) Grid topology for Section \ref{sec:Simulation}. The central node communicates with all nodes encircled by the solid red line (if $d=1$) or dashed blue line (if $d=2$).} \label{fig:grid_structure}
\end{minipage}
\vspace{-1em}
\end{figure}

\textbf{Structured communication.} We say that controller $\mathbf{K}$ obeys communication structure $\mathcal{S}^\text{comm} \in \{0, 1\}^{N \times N}$ if $\forall i,j$ s.t. $\mathcal{S}^\text{comm}_{ij} = 0$, $[\mathbf{K}(s)]_{ij} \equiv 0$.
To enforce \textbf{local communication}, it suffices to set $\mathcal{S}^\text{comm} = \mathcal{A}^d$ for some locality parameter $d \in \mathbb{Z}_+$, e.g., if $d=1$, then sub-controllers can only communicate if their subsystems are 1-hop neighbors on graph $\mathcal{G}$. 
An example is given in Fig. \ref{fig:grid_structure} (Right).

\textbf{Structured disturbance rejection.} We say that controller $\mathbf{K}$ achieves disturbance rejection with structure $\mathcal{S}^\text{dist} \in \{0, 1\}^{N \times N}$ if, in the closed-loop system (i.e., system \eqref{equ:LTIsys} with controller \eqref{eq:controller_time_domain}), disturbances $[w]_j$ can propagate to state $[x]_i$ only if $\mathcal{S}^\text{dist} \neq 0$. To enforce \textbf{local disturbance rejection}, it suffices to set $\mathcal{S}^\text{dist} = \mathcal{A}^d$, e.g., if $d=1$, then disturbances entering into one subsystem may only spread to another subsystem if they are 1-hop neighbors on graph $\mathcal{G}$.

We now present a clarifying example. Consider a plant with 3 subsystems, with state partition $[x]_1 = \begin{bmatrix}x_1 & x_2\end{bmatrix}^\top$, $[x]_2 = \begin{bmatrix}x_3 & x_4\end{bmatrix}^\top$, $[x]_3 = x_3$ and input partition $[u]_i = u_i$. Then, system matrices $A$ and $B$ can be partitioned as shown in \eqref{eq:example}; the corresponding graph $\mathcal{G}$ is shown in Fig. \ref{fig:grid_structure} (Left).
\begin{equation}
    A = \left[ \begin{array}{rr|rr|r}
        3 & 2 & 0 & 1 & 0 \\
        1 & 0 & 0 & 2 & 0 \\ \hline
        0 & 0 & 1 & 1 & 1 \\
        0 & 0 & 2 & 0 & 0 \\ \hline
        0 & 0 & 5 & 2 & 1
    \end{array}\right], \quad B = \left[ \begin{array}{r|r|r}
        0 & 0 & 0 \\
        1 & 0 & 0 \\ \hline
        0 & 0 & 0 \\
        0 & 1 & 0 \\ \hline
        0 & 0 & 1 \end{array}\right]
 \label{eq:example}
\end{equation}
\noindent The adjacency matrix for this system is: $\mathcal{A} = \begin{bmatrix} 1 & 1 & 0 \\ 0 & 1 & 1 \\ 0 & 1 & 1\end{bmatrix}$. 

\noindent If we set $\mathcal{S}^\text{comm} = \mathcal{A}$, then sub-controller 1 cannot receive communication from sub-controller 3, and sub-controllers 2 and 3 cannot receive communication from sub-controller 1. If we set $\mathcal{S}^\text{dist} = \mathcal{A}$, then disturbances in subsystem 1 may not propagate to any other subsystem; disturbances in subsystem 3 may not propagate to subsystem 1.

\subsection{Problem statement}
We seek a controller \eqref{eq:controller_time_domain} that minimizes some objective $f$ and satisfies the aforementioned structured constraints: 
\begin{equation} \label{eq:original_problem}
\begin{aligned}
    \min_\mathcal{K} \quad& f(x(t),u(t)) \\
    \text{s.t.} \quad & \dot{x}(t) = A x(t) + B u(t) + w(t), \quad u(t) = \mathcal{K}(x(0:t)) \\
    &\mathcal{K} \text{ internally stabilizing, linear, causal} \\
    & \mathcal{K} \text{ obeys communication structure } \mathcal{S}^\text{comm} \\
    & \mathcal{K} \text{ achieves dist. rejection with structure } \mathcal{S}^\text{dist}
\end{aligned}
\end{equation}
This problem is generally nonconvex due to requirements on both structured communication \textit{and} disturbance rejection. We use system level synthesis (SLS) to convexify it.

\section{System level synthesis} \label{sec:sls}
We review core elements of the SLS framework and begin translating them into the continuous-time setting. 
First, take the Laplace transform of \eqref{equ:LTIsys} and rearrange to get:
\begin{equation}
    (sI-A)\mathbf{x}(s) = B\mathbf{u}(s)+\mathbf{w}(s) \label{equ:LTIsys_s}
\end{equation}
Plugging in equation \eqref{equ:state_feedback} gives:
\begin{subequations}\label{equ:Phi} 
\begin{equation}
    \mathbf{x}(s) = (sI-A-B_{2}\mathbf{K}(s))^{-1}\mathbf{w}(s) =: 
    \Phibf_x(s)\mathbf{w}(s)
\end{equation}
\begin{equation}
    \mathbf{u}(s) = \mathbf{K}(s)\Phibf_x(s)\mathbf{w}(s) =: \Phibf_u(s) \mathbf{w}(s)
\end{equation}
\end{subequations}
where $\Phibf_x(s)$ and $\Phibf_u(s)$ are the \textit{closed-loop responses} from disturbance to state and control action, respectively. Theorem \ref{thm:feasibility} (analogous to Theorem 4.1 in \cite{anderson2019system}) shows that solving over stabilizing controllers ($\mathbf{K}$) is equivalent to solving over an affine subspace of closed-loop responses ($\Phibf_x, \Phibf_u$).

\begin{figure} 
 \begin{minipage}{0.65\linewidth}
 \centering
    \includegraphics[width=\linewidth, page=1]{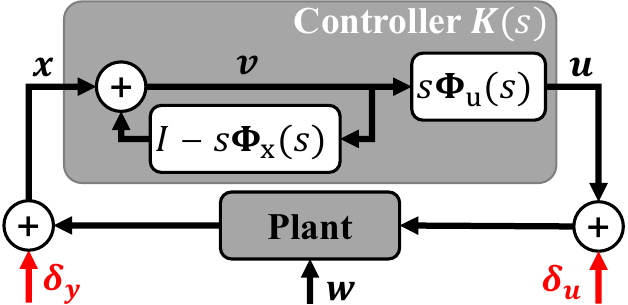}
\end{minipage}
\begin{minipage}{0.3\linewidth}
    \caption{Controller implementation \eqref{equ:controller_realization}, in closed loop with the plant and exogeneous disturbances/noise. Exogeneous signals $\boldsymbol{\delta}_u$ and $\boldsymbol{\delta}_y$ are used to verify internal stability of the system.} \label{fig:controller_structure}
\end{minipage}
\vspace{-1em}
\end{figure}

\begin{theorem}
    \label{thm:feasibility}
    For system \eqref{equ:LTIsys} evolving under causal, linear state feedback controller \eqref{equ:state_feedback}, the following hold:
    
    \textbf{Statement 1.} The affine subspace\footnote{This subspace is nontrivial if system $(A, B)$ is stabilizable.} defined by:
    \begin{subequations} \label{equ:affine_subspace}
        \begin{equation} \label{eq:feasibility1}
             \begin{bmatrix}
            sI-A & -B
        \end{bmatrix}
        \begin{bmatrix}
            \Phibf_x(s) \\ \Phibf_u(s)
        \end{bmatrix}
        = I
        \end{equation}
        \begin{equation} \label{eq:feasibility2}
            \Phibf_x(s), \Phibf_u(s) \in \RRational
        \end{equation}
    \end{subequations}
    parameterizes all closed-loop responses $\{\Phibf_x(s), \Phibf_u(s)\}$ that are achievable by an internally stabilizing controller $\mathbf{K}(s)$.
    
    \textbf{Statement 2.} $\forall \{\Phibf_x(s), \Phibf_u(s)\}$ satisfying \eqref{equ:affine_subspace}, the causal linear controller $\mathbf{K}(s) = \Phibf_u(s)\Phibf_x(s)^{-1}$, implemented in feedback configuration \eqref{equ:controller_realization} (see Fig. \ref{fig:controller_structure}) is internally stabilizing and achieves closed-loop responses $\{\Phibf_x(s), \Phibf_u(s)\}$.
    \begin{equation}
        \label{equ:controller_realization}
        \begin{split}
            \mathbf{v}(s) &= \mathbf{x}(s) + (I-s\Phibf_x(s))\mathbf{v}(s) \\
            \mathbf{u}(s) &= s\Phibf_u(s)\mathbf{v}(s)
        \end{split}
    \end{equation}
\end{theorem}

\begin{proof}
    \textit{(Statement 1)} Any stabilizing controller $\mathbf{K}(s)$ has associated closed-loop responses $\{\Phibf_x(s), \Phibf_u(s)\}$ that satisfy \eqref{equ:affine_subspace}; plug \eqref{equ:Phi} into \eqref{equ:affine_subspace} for the desired result.

    \textit{(Statements 1 \& 2)} We now show that if some transfer function matrices $\{\Phibf_x(s), \Phibf_u(s)\}$ satisfy \eqref{equ:affine_subspace}, there exists an internally stabilizing controller which achieves closed-loop responses $\{\Phibf_x(s), \Phibf_u(s)\}$. It suffices to show that the controller described by Statement 2 fulfils these requirements.

    First, we prove that $\Phibf_x(s)$ is invertible. We know that $\{\Phibf_x(s), \Phibf_u(s)\}$ satisfy \eqref{equ:affine_subspace}; rearranging \eqref{eq:feasibility1} yields:
    \begin{equation*}
        \Phibf_x(s) = (sI-A)^{-1}(I+B\Phibf_u(s))
    \end{equation*}  
    So $\Phibf_x(s)$ being invertible is equivalent to $I+B\Phibf_u(s)$ being invertible which, in turn, is equivalent to $I+B\Phibf_u(\infty)$ being invertible \cite{zhou1995robust}. Since $\Phibf_u(s)$ is strictly proper by \eqref{equ:affine_subspace}, $\Phibf_u(\infty) = 0$ and $I+B\Phibf_u(\infty) = I$, which is invertible. 
    
    Next, we show that this controller achieves the desired response. Plug $\mathbf{K}(s) = \Phibf_u(s)\Phibf_x(s)^{-1}$ into \eqref{equ:LTIsys_s} and \eqref{equ:state_feedback} and apply \eqref{eq:feasibility1} to yield the desired closed-loop responses $\mathbf{x}(s) = \Phibf_x(s) \mathbf{w}(s)$ and $\mathbf{u}(s) = \Phibf_u(s) \mathbf{w}(s)$. Finally, to show that this controller realization is internally stabilizing, consider perturbations $(\mathbf{w}, \boldsymbol{\delta}_u, \boldsymbol{\delta}_y)$ to the plant, control input, and plant output, respectively (see Fig. \ref{fig:controller_structure}). Build nine transfer functions from $(\mathbf{w}, \boldsymbol{\delta}_u, \boldsymbol{\delta}_y)$ to $(\mathbf{x}, \mathbf{u}, \mathbf{v})$ and verify that all are stable; follow the procedure from Section 4 of \cite{anderson2019system}.
\end{proof} 



In this parameterization, structured (and therefore local) communication and disturbance rejection can be achieved by enforcing sparsity constraints on closed-loop responses:
\begin{equation} \label{eq:sparse_constr}
    \Phibf_x(s) \in \mathcal{S}_x, \quad \Phibf_u(s) \in \mathcal{S}_u
\end{equation}
where $\mathcal{S}$ represents sparsity constraints encoding structured communication and disturbance rejection. With respect to the original problem \eqref{eq:original_problem}, $\mathcal{S} = \mathcal{S}^\text{comm} \cap \mathcal{S}^\text{dist}$\footnote{One limitation is that in this framework, it is impossible to enforce structured communication without enforcing structured disturbance rejection, and vice versa; see \cite{li2020separating} for a strategy that mitigates this}. 

Matrices $\{\Phibf_x(s), \Phibf_u(s)\}$ are also used to implement the controller via \eqref{equ:controller_realization}, which allows us to avoid inversion of $\Phibf_x$ and thus, preserve sparsity. We can then rewrite the original problem \eqref{eq:original_problem} in the SLS parameterization as:
\begin{equation} \label{equ:general_sls_problem}
\underset{\Phibf_x(s),\Phibf_u(s)}{\min} \hspace{0.5em}  g(\Phibf_x(s),\Phibf_u(s)) \hspace{0.5em} \text{s.t.} \hspace{0.5em} \eqref{equ:affine_subspace}, \eqref{eq:sparse_constr}
\end{equation}
where $g(\Phibf_x(s),\Phibf_u(s))$ is formulated to be equivalent to the original control objective (see Section \ref{sec:optimal_phi} for examples). If objective $g$ is convex, then this problem can be solved using convex optimization methods, since both constraints are affine with respect to the decision variables.

\textbf{Remark:} Enforcing sparsity constraints on the decision variables can also facilitate scalability if the objective and constraints are \textit{separable}. For example, if minimizing $g$ for each column of $\Phibf_x$ and $\Phibf_u$ is equivalent to minimizing $g$ over the full matrices, then instead of solving optimization \eqref{equ:general_sls_problem} directly, we can split it into smaller column-wise optimization problems be solved in parallel. If the decision variables are constrained to be sparse, the size of each column-wise problem will remain independent of size of the overall network. 
See \cite{anderson2019system, li2022distributed} for additional discussion.

\section{Closed-loop pole selection}
\label{sec:SPA_technique}

\subsection{Finite-dimensional truncation}
Problem \eqref{equ:general_sls_problem} involves solving over the infinite-dimensional affine subspace \eqref{equ:affine_subspace}. In previous work on discrete-time SLS, finite impulse response approximations are used to truncate this problem into a finite-dimensional one;  
we restrict ourselves to closed-loop responses that satisfy \eqref{equ:affine_subspace} \textit{and} place all closed-loop poles at zero, up to some fixed multiplicity. For continuous-time systems, an analogous approach is to first select some fixed, finite multiset $\mathcal{P}$ of closed-loop poles, and restrict ourselves to solving over all closed-loop responses that satisfy \eqref{equ:affine_subspace} and place closed-loop poles at $\mathcal{P}$. This is related to the notion of Ritz approximations in continuous time, which utilize a single pole with varying multiplicities \cite{linnemann1999convergent}, as well as simple pole approximation in discrete-time \cite{10347280}.

Given $\mathcal{P}$, some multiset of poles which are contained in the open left-half-plane (OLHP), we restrict ourselves to considering closed-loop responses of the form:
\begin{equation} \label{equ:phi_partial}
    \begin{aligned} 
    \Phibf_x(s) &= \sum_{l=1}^K\sum_{j=1}^{m_l}\Phi_x(l,j)\frac{1}{{(s-p_l)}^j} \\
    \Phibf_u(s) &= \sum_{l=1}^K\sum_{j=1}^{m_l}\Phi_u(l,j)\frac{1}{{(s-p_l)}^j}
    \end{aligned}
\end{equation}
where $K$ is the number of distinct poles in $\mathcal{P}$, and $m_l$ is the number of times $p_l$ appears in $\mathcal{P}$. We will refer to $\Phi_x(l,j)$ and $\Phi_u(l,j)$ as the \textit{spectral components} of $\Phibf_x(s)$ and $\Phibf_u(s)$, respectively. Now, we can rewrite constraint \eqref{equ:affine_subspace} in terms of the spectral components of the closed-loop responses: 

\begin{lemma} \label{lemm:CL_fixedpoles}
    Consider closed-loop responses of the form \eqref{equ:phi_partial}. If they are real-valued, then they lie in the affine subspace defined by \eqref{equ:affine_subspace} if and only if the following hold:
\begin{equation} \label{equ:general_constraints}
    \begin{split}
        1. \quad&\sum_{l=1}^K \Phi_x(l,1) = I \\
        2. \quad& \text{for } l \in \{1,2,\dots,K\} \text{ and } j \in \{1, 2,\dots, m_l-1\}: \\
        & \quad \Phi_x(l,j+1) + (p_lI-A)\Phi_x(l,j)-B\Phi_u(l,j) = 0  \\
        & \quad (p_lI-A)\Phi_x(l,m_l) -B\Phi_u(l,m_l) = 0 \\
    \end{split} 
\end{equation}
\end{lemma}
\begin{proof}
    \eqref{eq:feasibility2} holds by construction of $\mathcal{P}$ and \eqref{equ:phi_partial}. To show that \eqref{eq:feasibility1} holds, plug \eqref{equ:phi_partial} into \eqref{eq:feasibility1} and write out the constraint at each index (similar to \cite{Fisher2022ApproximationBS}).
\end{proof}

\textbf{Remark:} The number of poles required to satisfy \eqref{equ:general_constraints} does not depend on system size. For example, consider system \eqref{equ:LTIsys} with an arbitrary number of states and full rank $B$ matrix. It is sufficient to choose a single pole $p<0$; then, $\Phibf_x(s) = \frac{1}{s-p}I_n$ and $\Phibf_u(s) = \frac{1}{s-p}B^{-1}$ satisfy \eqref{equ:general_constraints}.

\textbf{Remark:} In general, spectral components $\Phi_x(l,j)$, $\Phi_u(l,j)$ need not be real; see Section \ref{sec:optimal_phi} for details.

We use simple pole approximation (SPA), where all poles have multiplicity 1 (i.e., $m_l = 1$ $\forall$ $l$). Then, \eqref{equ:phi_partial} becomes:
\begin{equation} \label{equ:phi_SPA}
    \Phibf_x(s) = \sum_{l=1}^K\Phi_x(l)\frac{1}{(s-p_l)}, \hspace{0.5em} 
    \Phibf_u(s) = \sum_{l=1}^K\Phi_u(l)\frac{1}{(s-p_l)}
\end{equation}
and constraint \eqref{equ:general_constraints} becomes:
\begin{equation} \label{equ:constraints_for_system_with_simple_poles}
    \sum_{l=1}^K \Phi_x(l) = I, \quad B\Phi_u(l) = (p_lI-A)\Phi_x(l)
\end{equation}
The remainder of this section focuses on SPA. However, optimization programs provided in Section \ref{sec:optimal_phi} can be used with poles from SPA or other methods (e.g., Ritz). We briefly compare SPA against other pole selection methods in Section \ref{sec:Simulation} and defer detailed exploration to future work.


\subsection{Simple pole approximation}
We now present SPA for continuous-time SLS. Our approach is inspired by \cite{10347280}; here, we devise a continuous-time version and rigorously demonstrate (Theorems \ref{thm:Lipschitz}-\ref{thm:SPA}, Lemma \ref{lemm:spa_continuous}) that poles generated via SPA indeed let us approximate the optimal continuous-time closed-loop responses.

Let $\Pcal$ now denote a set of $K$ poles generated by SPA, where $K$ is even. We first generate discrete-time poles $z_k \in \D$ using the Archimedes spiral and its reflection, then map them into the OLHP using the bilinear transform: 
\begin{equation}
    \begin{aligned}
        \text{for } k = 1 \ldots \frac{K}{2}, \quad  \theta_k = 2\sqrt{\pi k}, \quad r_k = \sqrt{\frac{k}{K+1}}  \\
        z_{k} = r_k e^{j\theta_k}, \quad
        z_{-k} = r_k e^{-j\theta_k}, \quad p_k = \frac{z_k-1}{z_k+1}
         \label{equ:pole_selection}
    \end{aligned}
\end{equation}

\noindent To show that this approximation method approaches the optimal solution, we first recall a result from prior work.

\begin{theorem}
    \label{thm:convergence_of_spiral} (Theorem 4 in \cite{10347280}) 
    Let $\Zcal_{\Pcal}$ be a set of $K$ poles generated by discrete-time SPA, i.e., $\Zcal_{\Pcal} := \{z_{k}\}_{k=-0.5K}^{0.5K}$. Let $\Zcal_{\Qcal}$ be some arbitrary set of poles in $\D$. Define $D(\Zcal_{\Pcal})$ as the max-min  distance between poles in $\Zcal_{\Pcal}$ and $\Zcal_{\Qcal}$, i.e.,
    \begin{equation}
        D(\Zcal_{\Pcal}) := \underset{x \in \Zcal_\Qcal}{\max}(\underset{y \in \Zcal_\Pcal}{\min}|x-y|) \\
    \end{equation}
    Then, $\underset{K\to+\infty}{\lim}D(\Zcal_{\Pcal}) \to 0$.
\end{theorem}

\noindent This theorem shows that SPA-generated poles can approximate the poles ($\Zcal_\Qcal$) of any discrete-time transfer function matrix. We begin to extend this to continuous-time: first, we show that poles close in the unit disk will also be close in the OLHP after applying the bilinear transform.

\begin{theorem}
    \label{thm:Lipschitz} Let $\Pcal$ and $\Qcal$ be two sets of poles in the OLHP. Consider $p \in \Pcal$, $q \in \Qcal$, $z_p = \frac{1+p}{1-p}$ and $z_q = \frac{1+q}{1-q}$. Define the following constants:
    \begin{align*}
    c_1 &= \frac{2c_4}{c_3(c_3+c_4)},  &\quad c_2 &= \frac{c_3(c_3+c_4)}{2} \\
    c_3 &= \max_{q \in \Qcal}|1-q|,    &\quad c_4 &= 1+\min_{q \in \Qcal}|q|
    \end{align*}
    If $|z_p-z_q| \leq c_1$, then the following must hold:
    \begin{equation*}
        |p-q| \leq c_2|z_p-z_q|, \quad |p-q| \leq c_4 
    \end{equation*}
\end{theorem}

\begin{proof}
    Since $p = \frac{z_p-1}{z_p+1}$ and $q = \frac{z_q-1}{z_q+1}$, we have:
    \begin{equation}
        |p - q| = \frac{2}{|1+z_p||1+z_q|}|z_p-z_q| \label{equ:transformed_p_q}
    \end{equation}
    One of the right-hand-side terms can be upper-bounded: 
    \begin{equation} \label{eq:thm3_step}
    \begin{split}
        \frac{2}{|1+z_p||1+z_q|} &= \frac{2}{|(1+z_q)+(z_p-z_q)||1+z_q|} \\
        & \leq \frac{2}{||1+z_q|-|z_p-z_q|||1+z_q|}
    \end{split}
    \end{equation}
    We can further refine this bound as follows: since $|1+z_q| = \frac{2}{|1-q|}$, we have $|1+z_q| \geq \frac{2}{{\max}_{q \in \Qcal}|1-q|} = \frac{2}{c_3}$. Since $|z_p-z_q| \leq c_1$, we have that $|z_p-z_q| \leq \frac{c_4}{c_4+c_3}\frac{2}{c_3} \leq \frac{c_4}{c_4+c_3}|1+z_q| < |1+z_q|$. Then, we can rewrite \eqref{eq:thm3_step} as:
    \begin{equation}
        \begin{aligned}
            \frac{2}{|1+z_p||1+z_q|}
            & \leq \frac{2}{(|1+z_q|-|z_p-z_q|)|1+z_q|} \\
            & \leq \frac{2}{(|1+z_q|-\frac{c_4}{c_4+c_3}|1+z_q|)|1+z_q|} \\
            & = \frac{2(c_3+c_4)}{c_3|1+z_q|^2} \leq c_2
        \end{aligned}
    \end{equation}
    Plugging this into \eqref{equ:transformed_p_q} yields $|p-q| \leq c_2|z_p-z_q|$. Combining this with $|z_p-z_q| \leq c_1$ yields $|p-q| \leq c_1c_2 = c_4$.
\end{proof}

This leads to a continuous-time analogue of Theorem \ref{thm:convergence_of_spiral}:

\begin{lemma} \label{lemm:spa_continuous}
    Let $\Pcal$ be a set of $K$ poles generated by continuous-time SPA as described in equation \eqref{equ:pole_selection}. Let $\Qcal$ be some arbitrary set of poles in the OLHP. Define $D(\Pcal)$ as the max-min distance between poles in $\Pcal$ and $\Qcal$, i.e.,
        \begin{equation} \label{eq:d_ct}
        D(\Pcal) = \underset{x \in \Qcal}{\max}(\underset{y \in \Pcal}{\min}|x-y|)
        \end{equation}
    It is always possible to select $K$ such that $D(\Pcal) \leq c_4$.
\end{lemma}
\begin{proof}
    Let $\Zcal_\Pcal$ and $\Zcal_\Qcal$ represent the bilinear transforms of sets $\Pcal$ and $\Qcal$. For any $z_p \in \Zcal_\Pcal$ and $z_q \in \Zcal_\Qcal$, $|z_p - z_q| \leq     D(\Zcal_{\Pcal})$ by definition. From Theorem \eqref{thm:convergence_of_spiral}, we can always choose $K$ to make $D(\Zcal_{\Pcal})$ arbitrarily small; thus, we can always choose $K$ so that $|z_p - z_q| \leq c_1$. Applying Theorem \eqref{thm:Lipschitz} gives the desired result.
\end{proof}

Now, the final result: transfer function matrices using SPA-generated poles can approximate arbitrary stable transfer function matrices in continuous time. We write transfer function matrices with SPA-generated poles as:
     \begin{equation}
         \mathbf{G}(s) = \sum_{p \in \Pcal}G_p\frac{1}{s-p}
     \end{equation}
where spectral components $\{G_p\}_{p \in \Pcal}$ are constant matrices.

\begin{theorem}
     \label{thm:SPA}
     Let $\Pcal$ be a set of poles generated by continuous-time SPA per \eqref{equ:pole_selection}; select $K$ so that $D(\Pcal) \leq c_4$. Then, for any transfer function matrix $\mathbf{S}(s) \in \frac{1}{s}\mathcal{RH}_\infty$, there exists constants $c_s, c_s' > 0$ and matrices $\{G_p\}_{p \in \Pcal}$ such that
     \begin{subequations} \label{eq:thm_norm_bound}
         \begin{equation}
             \norm{\sum_{p \in \Pcal}G_p\frac{1}{s-p}-\mathbf{S}(s)}_{\mathcal{H}_2} \leq {c_s}D(\Pcal) 
         \end{equation}
         \begin{equation}
             \norm{\sum_{p \in \Pcal}G_p\frac{1}{s-p}-\mathbf{S}(s)}_{\mathcal{H}_\infty} \leq {c_s'}D(\Pcal) 
         \end{equation}
      \end{subequations}
\end{theorem}

This result and its proof are analogous to Theorem 1 in \cite{10347280}; a proof sketch is provided in the
\ifarxivversion
Appendix.
\else
Appendix \footnote{Available at \url{https://arxiv.org/abs/2410.08135}}.
\fi

\section{\texorpdfstring{$\H2$}{} and \texorpdfstring{$\Hinf$}{} Control via SLS}
\label{sec:optimal_phi}
SLS problem \eqref{equ:general_sls_problem} is infinite-dimensional. To bypass this, we constrain closed-loop responses $\Phibf_x$ and $\Phibf_u$ to have pre-generated stable poles, and optimize over spectral components. These results can be used with poles generated through SPA or other methods (e.g., Ritz).

To simplify notation, we re-order the SPA-generated pole set $\Pcal$: $\Pcal = \{ p_1, p_{-1},\dots p_{\frac{K}{2}},p_{-\frac{K}{2}}\}$. For the remainder of the paper (except the Appendix), we re-label these poles as $\Pcal = \{ p_1, p_2,\dots p_K\}$, with the convention that $(p_{2i-1}, p_{2i})$ is a complex conjugate pair. While poles can be complex, transfer function matrices $\Phibf_x(s)$ and $\Phibf_u(s)$ must be real-valued; thus, spectral components must satisfy the following:
\begin{equation}
\begin{aligned}
    \quad \forall i \in \{1,2,\ldots, \frac{K}{2}\}, \quad \Phi_x(2i-1) &= \overline{\Phi_x(2i)}, \\
    \Phi_u(2i-1) &= \overline{\Phi_u(2i)},  
    \label{equ:constraints_to_be_real}
\end{aligned}
\end{equation} 

Then, for poles chosen via SPA, feasibility constraints \eqref{equ:affine_subspace} can be enforced via \eqref{equ:constraints_to_be_real} and \eqref{equ:constraints_for_system_with_simple_poles}. Similarly, sparsity constraints \eqref{eq:sparse_constr}  can be enforced via:
\begin{equation}
    \Phi_x(l) \in  \mathcal{S}_x,  \quad \Phi_u(l) \in  \mathcal{S}_u  \label{equ:sparsity_constraints}
\end{equation}

The SLS-based control problem \eqref{equ:general_sls_problem} now becomes: 
\begin{equation} \label{eq:sls_spec_component}
\begin{aligned}
    \min_{\Phi_x(l),\ \Phi_u(l), \text{ } l = 1 \ldots K}\quad & {g(\Phibf_x(s),\Phibf_u(s))}  \\
    \text{subject to} \quad & \eqref{equ:phi_SPA}, \eqref{equ:constraints_to_be_real}, \eqref{equ:constraints_for_system_with_simple_poles}, \eqref{equ:sparsity_constraints}
\end{aligned}
\end{equation}
where we optimize over spectral components. We now derive expressions for objective $g$ in terms of spectral components.

\subsection{State-space realization}
We consider linear quadratic (LQ) cost formulations with state penalty matrix 
$Q\succeq 0$ and input penalty matrix $R\succ 0$. We introduce augmented variable $\Psibf(s)$:
\begin{equation}
    \Psibf(s) :=\begin{bmatrix}
    Q^{\frac{1}{2}} & 0 \\ 0 & R^{\frac{1}{2}}
\end{bmatrix}\begin{bmatrix}
    \Phibf_x(s) \\ \Phibf_u(s)
\end{bmatrix}
\label{equ:Psi_s}
\end{equation}
We can write the spectral components of $\Psibf(s)$ as linear combinations of spectral components of $\Phibf_x(s)$ and $\Phibf_u(s)$:
\begin{equation} \label{equ:psi_definition}
    \begin{aligned}
    \Psibf(s) &= \sum_{l=1}^K\Psi(l)\frac{1}{s-p_l} \\
    \Psi(l) &=  \begin{bmatrix}
    Q^{\frac{1}{2}} & 0 \\ 0 & R^{\frac{1}{2}}
    \end{bmatrix}
    \begin{bmatrix}
        \Phi_x(l) \\ \Phi_u(l)
    \end{bmatrix} 
\end{aligned}
\end{equation}

The objective is to minimize $\|\Psibf(s)\|_{\H2}$ for $\H2$ control, and $\|\Psibf(s)\|_{\Hinf}$ for $\Hinf$ control. To facilitate this, we use the following state-space realization of \eqref{equ:psi_definition}:
\begin{align} \label{equ:ABC_structure_for_ss_realization} \nonumber
    \dot{x} &= \mathcal{A}x+\mathcal{B}u =: \begin{bmatrix}
        p_1 I_n & & \\ & \ddots & \\ & & p_K I_n
    \end{bmatrix}x + \begin{bmatrix}
        I_n \\ \vdots \\ I_n
    \end{bmatrix}u \\
    y &= \mathcal{C}x =: \begin{bmatrix}
        \Psi(1) & \ldots & \Psi(K)
    \end{bmatrix}x
\end{align}

When using the SPA-generated poles, all poles are complex, which means realization \eqref{equ:ABC_structure_for_ss_realization} is complex. However, if \eqref{equ:constraints_to_be_real} is satisfied, then we can construct orthogonal matrix
\begin{align}
    T = \text{blkdiag}(U, U, \ldots U), \quad U = \begin{bmatrix}
            \frac{1+j}{2} &  \frac{1-j}{2} \\ \frac{1-j}{2} & \frac{1+j}{2}
        \end{bmatrix}
\end{align}
such that the similarity-transformed matrices 
\begin{equation}
    \tilde{\mathcal{A}} = T^*\mathcal{A}T, \quad \tilde{\mathcal{B}} = T^*\mathcal{B}, \quad \tilde{\mathcal{C}} = \mathcal{C}T \label{equ:transformed_ABC}
\end{equation}
are real-valued. Even when poles are not generated by SPA, as long as $\Psibf(s)$ is real-valued, we can always find a real-valued state-space realization for the controller. For general non-repeated poles, partition $\mathcal{A}$ into $\mathcal{A} = \text{blkdiag}(\mathcal{A}_c, \mathcal{A}_r)$, where $\mathcal{A}_c$ contains complex conjugate poles (where adjacent values are conjugate pairs, similar to SPA) and $\mathcal{A}_r$ contains real-valued poles. Then, apply $T = \text{blkdiag}(U, U, \ldots U, I)$. For repeated poles (e.g., Ritz), let the  $(i,j)^\text{th}$ element of transfer function matrix $\Psibf(s)$ be written as
\begin{equation}
    \Psibf_{ij}(s) = \sum_{l=1}^K\frac{\Psi_{ij}(l)}{(s-p)^l}
\end{equation}
A state-space realization for $\Psibf_{ij}(s)$ is:
\begin{equation}
 \label{eq:partitioned_realizations}
\begin{aligned}
    \mathcal{A}^{ij} &= \begin{bmatrix}
    p & 1 & 0 & \cdots & 0 \\
    0 & p & 1 & \cdots & 0 \\
    0 & 0 & p & \cdots & 0 \\
    \vdots & \vdots & \vdots & & \vdots \\
    0 & 0 & 0 & \cdots & p
\end{bmatrix}, \quad \mathcal{B}^{ij} =  \begin{bmatrix}
    1 \\ 1 \\ \vdots \\ 1
\end{bmatrix}, \\
\mathcal{C}^{ij} &= \begin{bmatrix} \Psi_{ij}(K) & \mathcal{C}^{ij}_{K-1} \ldots \mathcal{C}^{ij}_1 \end{bmatrix}, \text{where } \\
&\mathcal{C}^{ij}_k = \Psi_{ij}(k)-\Psi_{ij}(k+1) \text{ for } k=1 \ldots K-1
\end{aligned}
\end{equation}
where $\mathcal{A}^{ij} \in \mathbb{R}^{K \times K}$. 
Concatenate state-space realizations $(\mathcal{A}^{ij}, \mathcal{B}^{ij},\mathcal{C}^{ij})$ to produce a realization for $\Psibf(s)$.

The remainder of this section computes $\H2$ and $\Hinf$ objectives using these state-space realizations.
We use ($\tilde{\mathcal{A}}, \tilde{\mathcal{B}}, \tilde{\mathcal{C}}$) to indicate the appropriate real-valued state-space realization.

\subsection{\texorpdfstring{$\mathcal{H}_2$}{H2} control}
From linear matrix inequality (LMI) results in \cite{599969}, we know that $\norm{\Psibf(s)}_{\H2}<\eta$ if $\exists P \succ 0$, $W \succ 0$ such that
\begin{equation}
    \begin{aligned}
        \begin{bmatrix}\tilde{\Acal}^{\top}P+P\tilde{\Acal} & P\tilde{\Bcal} \\ \tilde{\Bcal}^{\top}P & -I
        \end{bmatrix} \prec 0, \quad  
        \begin{bmatrix}
            P & \tilde{\Ccal}^{\top} \\ \tilde{\Ccal} & W
        \end{bmatrix} \succ 0, \quad
        \text{trace}(W) < \eta
    \end{aligned}
    \label{equ:H2_lmi}
\end{equation}

Then, we can write the $\mathcal{H}_2$ problem as:
\begin{equation} 
\begin{aligned}
    \min_{\substack{
    \Phi_x(l),\ \Phi_u(l), \text{ } l=1 \ldots K \\ 
    P \succ 0,\ W\succ 0,\ \eta \geq 0
    }}{\quad \eta}  \\
    \text{subject to} \quad & \eqref{equ:phi_SPA}, \eqref{equ:constraints_to_be_real}, \eqref{equ:constraints_for_system_with_simple_poles}, \eqref{equ:sparsity_constraints}, \\ 
    & \eqref{equ:psi_definition},
    \eqref{equ:ABC_structure_for_ss_realization},
    \eqref{equ:transformed_ABC}, 
    \eqref{equ:H2_lmi} 
\end{aligned}
\end{equation}
which is convex. Constraint \eqref{equ:H2_lmi} can be partitioned into smaller parallel constraints by using partitioned matrices (similar idea as \eqref{eq:partitioned_realizations}); then, this problem is separable and therefore scalable (see Section \ref{sec:sls}). When poles are distinct (e.g., generated via SPA), we can equivalently write:
\begin{equation}
\begin{aligned}
    \min_{\Phi_x(l),\ \Phi_u(l), \text{ } l = 1 \ldots K}\quad & {\sum_{i=1}^m\sum_{j=1}^n R_{ij}}  \\
    \text{subject to} \quad & \eqref{equ:phi_SPA}, \eqref{equ:constraints_to_be_real}, \eqref{equ:constraints_for_system_with_simple_poles}, \eqref{equ:sparsity_constraints}, \eqref{equ:psi_definition} \\
    &  R_{ij} = \sum_{l=1}^K\lim_{s \to p_l}(s-p_l)\overline{\Psi_{ij}(s)}\Psi_{ij}(s)
\end{aligned}
\end{equation}
\vspace{-3em}

\subsection{\texorpdfstring{$\mathcal{H}_\infty$}{Hinf} control}
First, recall the KYP lemma (Lemma 4.1 in \cite{gahinet1994linear}):
\begin{lemma}
    \label{lem:KYP}
    Consider a strictly proper transfer function $G(s)$ with real-valued realization $(\tilde{\mathcal{A}}, \tilde{\mathcal{B}}, \tilde{\mathcal{C}}, 0)$. Then, the statement
    \begin{equation}
        \norm{\tilde{\mathcal{C}}{(sI-\tilde{\mathcal{A}})}^{-1}\tilde{\mathcal{B}}}_{\Hinf} < \gamma \text{ and } \tilde{\mathcal{A}} \text{ is Hurwitz }
    \end{equation}
    is equivalent to the statement
    \begin{equation} \label{eq:kyp}    
                 \exists X\succ 0 \text{ s.t.}
                 \begin{bmatrix}{\tilde{\mathcal{A}}^\top}X+X\tilde{\mathcal{A}} & X\tilde{\mathcal{B}} & \tilde{\mathcal{C}}^\top \\ 
        {\tilde{\mathcal{B}}^\top}X & -\gamma I & 0 \\
            \tilde{\mathcal{C}} & 0 & -\gamma I
    \end{bmatrix} \prec 0
    \end{equation}
\end{lemma}
Using this, we can write the convex $\mathcal{H}_\infty$ program:
\vspace{-0.5em}
\begin{equation} 
\begin{aligned}
    \min_{\substack{
    \Phi_x(l),\ \Phi_u(l), \text{ } l=1 \ldots K \\ 
    X \succ 0,\ \gamma \geq 0
    }}{\quad \eta}  \\
    \text{subject to} \quad & \eqref{equ:phi_SPA}, \eqref{equ:constraints_to_be_real}, \eqref{equ:constraints_for_system_with_simple_poles}, \eqref{equ:sparsity_constraints}, \\ 
    & \eqref{equ:psi_definition},
    \eqref{equ:ABC_structure_for_ss_realization},
    \eqref{equ:transformed_ABC}, 
    \eqref{eq:kyp} 
\end{aligned}
\end{equation}

\section{Simulations} \label{sec:Simulation}
We simulate our controller synthesis technique on two plants, setting $Q$ and $R$ as identity matrices. Simulations can be reproduced using the SLS-MATLAB toolbox \cite{li2019sls}.   

\subsection{Chain plant}
\begin{figure}
    \centering
    \includegraphics[width=0.8\linewidth]{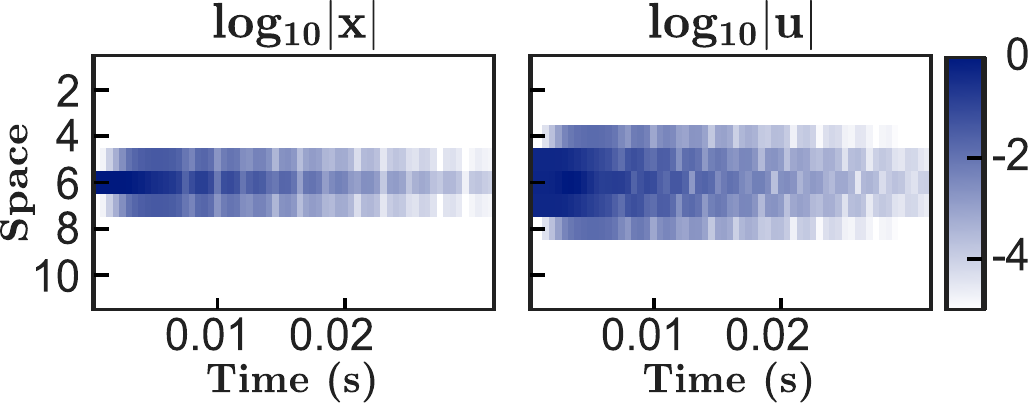}
    \vspace{-0.5em}
    \caption{Chain plant with SLS controller.  
    A disturbance is injected at node 6 at time $t=0$; it propagates only to nodes 4, 5, 6, 7, and 8 (i.e., nodes within $d=2$ distance of node 6) before being rejected.}
    \label{fig:sim_d2}
\vspace{-1em}
\end{figure}

This unstable plant has 11 nodes; $A$ is tri-diagonal, with $A_{ii}=0.6$, $A_{i,i-1} = A_{i-1,i} = 0.4$, and $B=I$. We use a distributed $\H2$ controller with 4 SPA-generated poles and locality parameter $d=2$.
The $\H2$ cost of our controller is $1.18$ times the cost of the centralized LQR controller. By design, this controller achieves local communication and local disturbance rejection (see Fig. \ref{fig:sim_d2}).

\subsection{Grid plant}
We consider linearized swing equations, as is done in \cite{anderson2019system}. The state of node $i$ is $[x]_i := {[\theta_i \quad \dot{\theta}_i]}^\top$, with dynamics
\begin{equation}
\begin{aligned}
    \left[ \dot{x} \right]_i =& \begin{bmatrix}
            0 & 1 \\ \frac{-\sum_{j \in \mathcal{N}_i}k_{ij}}{m_i} & \frac{-\beta_i}{m_i}
        \end{bmatrix}[x]_i + \sum_{j \in \mathcal{N}_i}\begin{bmatrix}
            0 & 0 \\  \frac{k_{ij}}{m_i} & 0
        \end{bmatrix}[x]_j \\
        & + \begin{bmatrix}
            0 \\ 1
        \end{bmatrix}u_i + \begin{bmatrix}
            0 \\ 1
        \end{bmatrix}w_i
\end{aligned}
\end{equation}
where parameters $m_i^{-1}$, $\beta_i$, and $k_{ij}$ represent inertia, damping, and coupling, respectively; $\theta_i$, $w_i$, and $u_i$ are phase angle deviation, disturbance, and control action, respectively; $\mathcal{N}_i$ indicates neighbors of node $i$. We generate a random grid topology (Fig. \ref{fig:grid_structure}) and random parameters using similar distributions as in \cite{anderson2019system}; the resulting plant is marginally stable.

\begin{figure}
    \centering
    \includegraphics[width=\linewidth]{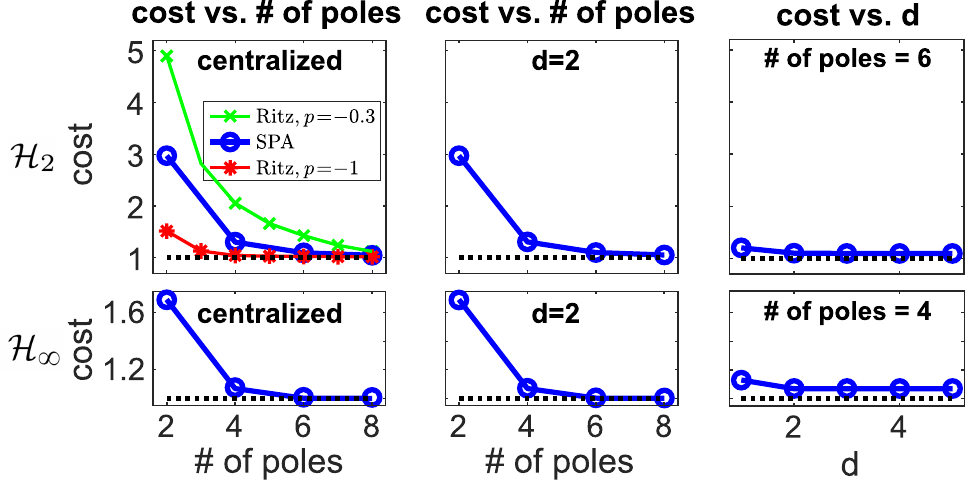}
    \vspace{-1.5em}
    \caption{Grid plant with various SLS controllers. All costs are normalized by the cost of the optimal centralized controller. For this system, $d=5$ is equivalent to a centralized controller since the diameter of the graph is 5.}
    \label{fig:h2hinf_sims}
\vspace{-1.5em}    
\end{figure}

We design $\H2$ and $\Hinf$ controllers with varying locality parameters ($d$) and number of poles. Results are shown in Fig. \ref{fig:h2hinf_sims}. As $d$ or the number of poles decreases, performance becomes less optimal; the $\Hinf$ controller performance is less sensitive in this regard than $\H2$. Note that near-optimal $\H2$ and $\Hinf$ controllers can be designed via SPA using only 6 poles, even though the true optimal closed-loop has 18 poles. We also include results using Ritz approximations for $\H2$, varying the multiplicity of the pole instead of the number of poles; we find that this method is impractical as it is quite sensitive to the choice of pole placement. 

We also compared our continuous-time controllers to discrete-time SLS controllers with zero-order hold. We let each controller have the same number of free variables, using  horizon $T=7$ for discrete-time and 6 SPA-generated poles for continuous-time. The resulting $\H2$ costs are 5.22 for the continuous-time case, and 5.48, 5.32, and 6.98 for the discrete-time case corresponding to sampling time $\Delta t = 0.05$, $0.1$, and $0.2$ seconds. The continuous-time controller outperforms all discrete-time controllers; as expected, there is benefit in designing directly in the continuous-time domain. 

\bibliography{references}
\bibliographystyle{IEEEtran}

\ifarxivversion
\section{Appendix: Proof sketch for Theorem 4} \label{sec:appendix}

To prove this theorem, we first require Lemmas  \ref{lem:approximating_a_pole}-\ref{lem:bound_H2_with_Hinf} and Corollary \ref{cor:Hinf_to_d}.

   \begin{lemma}
       \label{lem:approximating_a_pole}
       $\forall m \in \mathbb{N}_+$, let $p_1, \cdots, p_m, q \in \mathbb{C}$ and $Re(q) \neq 0$. Then $\exists$ constants $c_1, \cdots, c_m$ s.t. when $Re(s) = 0$,
       \begin{align*}
           & |{\sum_{i=1}^m\frac{c_i}{s-p_i}-\frac{1}{(s-q)^m}}| \leq \\
           & \frac{\gamma^m}{\prod_{i-1}^m |Re(p_i)|}\sum_{k=1}^m\sum_{\substack{\mathcal{S} \subset I_m \\|\mathcal{S}| = k}}\sum_{i=1}^k\sum_{\substack{\mathcal{T} \subset \mathcal{S} \\|T| = i}}{|q|}^{k-i}\prod_{j \in \mathcal{T}}\vert p_j-q \vert
       \end{align*}
       where $\gamma = {max}_{q \in \Qcal}\{Re(q), \frac{1}{Re(q)}\}$
   \end{lemma}
   \begin{proof}
       Follows from algebraic manipulation of Lemma 2 in \cite{10347280}.
   \end{proof}

   \begin{corollary}
       \label{cor:Hinf_to_d}
       $\forall m \in \mathbb{N}_{+}$, let $p_1, \cdots, p_m, q \in \mathbb{C}$ and $Re(q) < 0$. Then, $\exists c_1 \cdots c_m$ s.t.
       \begin{equation}
           {\Vert \sum_{i=1}^m \frac{c_i}{s-p_i} - \frac{1}{(s-q)^m} \Vert}_{\mathcal{H}_\infty} \leq \frac{(2^m+1)(|q|+1)^{m-1}}{(r/\gamma)^m}\hat{d}(q)
       \end{equation}
       \label{fundamental_error_bound}
   \end{corollary}

   \begin{lemma}
       \label{lem:bound_H2_with_Hinf} 
       For SISO $G(s) \in \frac{1}{s}\mathcal{RH}_\infty$, there exists constant $\epsilon$ s.t.
       \begin{equation}
           {\Vert G(s) \Vert}_{\mathcal{H}_2} \leq \epsilon {\Vert G(s) \Vert}_{\mathcal{H}_\infty} \label{equ:H2_by_Hinf}
       \end{equation}
       \begin{proof}
           Follows from application of results in \cite{DEBRUYNE1995173})
       \end{proof}

       Assume that $G(s)$ has $n$ poles denoted by $p_1, \cdots, p_n$, $m$ zeros denoted by $z_1,\cdots, z_m$ ($m > n$) and DC gain $k$. Then, a possible $\epsilon$ is
       \begin{equation}
           \epsilon = \frac{1}{\sqrt{2|Re(p_n)|}} \prod_{j=1}^m M_j\prod_{j=m+1}^{n-1} \frac{1}{|Re(p_j)|}
       \end{equation}
       where
       \begin{equation}
           M_j = \sqrt{1+\frac{|Im(z_j)-Im(p_j)|}{\vert{|p_j|-Im(p_j)}\vert}+\frac{\vert{|p_j|-|z_j|}\vert}{{Re(p_j)}^2}}
       \end{equation}
   \end{lemma}

   Now, the proof sketch for Theorem \eqref{thm:SPA} is given. Let $\Qcal$ denote the poles of $\mathbf{S}(s)$; let $\Qcal_R$ denote its real poles $\Qcal_C$ denote its remaining poles. Using $m_q$ to denote the multiplicity of pole $q$, $\mathbf{S}(s)$ can be expressed in the following partial fraction form
   \begin{equation}
         \mathbf{S}(s) = \sum_{q \in \Qcal_R}\sum_{j=1}^{m_q}S_{(q,j)}\frac{1}{{(s-q)}^j} + \sum_{q \in \Qcal_C}\sum_{j=1}^{m_q}S_{(q,j)}\frac{1}{{(s-q)}^j}
   \end{equation}

   $\forall q \in \Qcal$, $j \in I_{m_q}$. Let $\Pcal(q,j) \subset \Pcal(q)$ denote the $j$ closest poles in $\Pcal$ to $q$. To approximate the pole $q$ with multiplicity $j$ using poles in $\Pcal(q,j)$, the coefficient $\{c_p^{(q,j)}\}_{p \in \Pcal(q,j)}$ is selected according to Lemma \ref{lem:approximating_a_pole}. To approximate the pole $\overline{q}$ with multiplicity $j$ using poles in $\Pcal(\overline{q},j)$, the coefficient $\{c_p^{(\overline{q},j)}\}_{p \in \Pcal(\overline{q},j)}$ is selected according to Lemma \ref{lem:approximating_a_pole}. It can be seen that $\{c_p^{(\overline{q},j)}\}_{p \in \Pcal(\overline{q},j)} = \overline{\{c_p^{(q,j)}\}_{p \in \Pcal(q,j)}}$. Also, since $\Pcal$ is closed under complex conjugation, we may select $\Pcal(\overline{q},j) = \overline{\Pcal(q,j)}$. Note that if $q$ is real, there will be two sets of poles $\Pcal(q,j)$ and $\overline{\Pcal(q,j)}$ to approximate $q$. Using Corollary \ref{fundamental_error_bound}, the following holds:
   \begin{equation}
       \begin{split}
           \norm{\sum_{p \in \Pcal}\frac{c_p^{(q,j)}}{s-p} - \frac{1}{{(s-q)}^j}}_{\mathcal{H}_\infty} & \leq \frac{(2^j+1)(|q|+1)^{j-1}}{(r/\gamma)^j}D(\Pcal)
       \end{split}
   \end{equation}
   Then, we can write out the spectral components of approximating transfer function $\mathbf{G}(s)$ as
   \begin{equation}
       G_p = \sum_{q \in \Qcal_R}\sum_{j=1}^{m_q}\frac{c_p^{(q,j)}+c_p^{(\overline{q},j)}}{2}S_{(q,j)}+\sum_{q \in \Qcal_C}\sum_{j=1}^{m_q}c_p^{(q,j)}S_{(q,j)}
   \end{equation}
   We can show that $\mathbf{G}(s)$ is real using the properties of $\Pcal$ and previously written relationships between coefficients. Then, after some algebraic simplifications, we find that 
   \begin{equation}
       c_s' = \sum_{q \in \Qcal} \sum_{j=1}^{m_q}{\Vert S_{(q,j)}\Vert}_2\frac{(2^j+1)(|q|+1)^{j-1}}{(r/\gamma)^j}
   \end{equation}
    To obtain $c_s$ involves algebraic manipulations and application of Lemma \eqref{lem:bound_H2_with_Hinf}, which ends in the finding that
    \begin{equation}
        c_s \coloneq \sum_{q \in \Qcal} \sum_{j=1}^{m_q}{\epsilon^{(q,j)}\Vert S_{(q,j)}\Vert}_2\frac{(2^j+1)(|q|+1)^{j-1}}{(r/\gamma)^j}
    \end{equation}
\fi

\end{document}